\documentstyle[preprint,aps]{revtex}

\begin{document}
\draft
\preprint{RU9562}
\title{Critical point multiplicities and multiplicity fluctuations in
heavy ion collisions}
\author{K. C. Chase and A. Z. Mekjian}
\address{Department of Physics, Rutgers University \\
Piscataway, New Jersey 08855 USA}
\date{\today}

\maketitle

\begin{abstract}
An exactly solvable model of nuclear fragmentation is shown to lead to
a simple connection between the critical point multiplicity
$\langle m \rangle_{c}$
and the critical point exponent $\tau$ recently reported on in the EOS
collaboration.  The importance of multiplicity fluctuations on
critical point behavior is also discussed.
\end{abstract}

\narrowtext


In a recent paper~\cite{Gilkes1994} the EOS collaboration reported
some very interesting results on the critical point behavior of
nuclear matter.  The analysis of the data is based on percolation
model predictions due to Campi~\cite{Campi1986}.
In this analysis, three parameters
appear, one being the expected critical multiplicity $\langle m
\rangle_{c}$ and the other
two critical exponents $\gamma$ and $\beta$.  The $\gamma$ and $\beta$
exponents are then used to obtain the critical exponent $\tau$.  The
purpose of this paper is to show there is a simple connection between
the critical exponent $\tau$ and the critical multiplicity
$\langle m \rangle_{c}$.
The development of this
relation is based on an exactly solvable statistical model of hadronic
matter which shares some features with percolation models.  Namely, in
the infinite mass number $A$ limit, an infinite cluster can appear below
a critical point in a variable $x$ called the tuning parameter, just
as in percolation models an infinite cluster may exist only at and above
a critical site or bond probability $p_{c}$.
Thermodynamic and statistical arguments relate the tuning parameter
$x$ to the volume, temperature, binding energy coefficient $a_{V}$,
and level spacing parameter $\varepsilon_{0}$,
namely
\begin{equation}
x = {V \over v_{0}(T)} \exp \left\{-{a_{V} \over k_{B}T} - {k_{B}T \over
\varepsilon_{0}} {T_{0} \over T+T_{0}} \right\} \;.
\end{equation}
Further details can be found in Ref.~\cite{Chase1994}.

A parallel of the exactly solvable model of hadronic matter with Bose
condensation and Feynman's approach to the $\lambda$ transition in
liquid helium was also noted in previous studies~\cite{Chase1994}.
In this parallel, the cluster size in fragmentation is equivalent to
the cycle length in the cycle class decomposition of the symmetric
group for the Bose problem.  The number of clusters of size $k$ is
the same as the number of cycles of length $k$.  The complete
fragmentation into nucleons corresponds to all unit cycles, which is
the identity permutation.  Bose condensation corresponds to the
appearance of longer and longer cycles in the cycle class
decomposition of the symmetric group.

In the statistical model of Ref.~\cite{Chase1994}
each fragmentation outcome happens
with a probability proportional to the Gibbs weight
\begin{equation}
P(\vec{n}) = {1 \over Z_{A}(\vec{x})}
  \prod_{k \ge 1} {x_{k}^{n_{k}} \over n_{k}!}
\end{equation}
where $Z_{A}(\vec{x})$ is the canonical partition function.
Here $x_{k} = x/k^{\tau}$ with $x$ as given above and
$\tau$ the critical exponent
originally introduced in nuclear fragmentation by Finn, et
al.\cite{Finn1982}.
Cluster distributions can be obtained from the partition functions
$Z_{A}(\vec{x})$ using
\begin{equation}
\langle n_{k} \rangle
  = x_{k} {Z_{A-k}(\vec{x}) \over Z_{A}(\vec{x})}
\end{equation}
In turn the $Z_{A}(\vec{x})$ can be obtained by recursive techniques,
and in particular
\begin{equation}
Z_{A}(\vec{x})
  = {1 \over A} \sum_{k \ge 1} k x_{k} Z_{A-k}(\vec{x})
\end{equation}
with $Z_{0}(\vec{x}) = 1$.  If $A$ is very large we can work in
the grand canonical limit, where
$\langle n_{k} \rangle = x z^{k}/k^{\tau}$,
$\langle m \rangle = x g_{\tau}(z)$ and $A = x g_{\tau-1}(z)$.
Here $z \le 1$ is the fugacity and $g_{\tau}(z) = \sum_{k} z^{k}/k^{\tau}$.
At $z = 1$, $g_{\tau-1}(z)$ is finite only if $\tau > 2$, which indicates
a critical point at $A = x_{c} g_{\tau-1}(1)$.
At this point the expected multiplicity $\langle m \rangle_{c}$
is given by
\begin{equation}
{\langle m \rangle_{c} \over A} = {\zeta(\tau) \over \zeta(\tau-1)}
\end{equation}
where $\zeta(x)$ is the Riemann zeta function.

The EOS collaboration recently reported independent determinations of
$\langle m \rangle_{c}$ and $\tau$.  Specifically, they found
$\langle m \rangle_{c} = 26 \pm 1$ and $\tau = 2.14 \pm 0.06$
for the fragmentation of gold nuclei by using a percolation model analysis.
For our model, $\langle m \rangle_{c} = 26 \pm 1$ implies
a critical exponent $\tau = 2.262 \pm 0.013$ which is consistent with
EOS result, but based on a different model.

The critical behavior of this model is most clearly seen from the
behavior of the multiplicity fluctuations
$\langle m^{2} \rangle - \langle m \rangle^{2} \equiv \langle m \rangle_{2}$.
In the grand canonical limit, these fluctuations are given by
\begin{equation}
\langle m \rangle_{2} = \left\{
  \begin{array}{ll}
    \langle m \rangle & \langle m \rangle \le \langle m \rangle_{c} \\
    \langle m \rangle - x {g_{\tau-1}(z)^{2} \over g_{\tau-2}(z)}
    & \langle m \rangle > \langle m \rangle_{c}
  \end{array} \right. \;,
\end{equation}
which is continuous function with a maximum at the transition point.
However the slope of $\langle m \rangle_{2}$ is discontinuous at the
transition point,
i.e. the peak at the critical point is a cusp.  Specifically,
\begin{equation}
\left. x {\partial \over \partial x} \langle m \rangle_{2}
  \right|_{x \rightarrow x_{c}^{+}} -
\left. x {\partial \over \partial x} \langle m \rangle_{2}
  \right|_{x \rightarrow x_{c}^{-}} = -2 A x_{c}
\end{equation}
Such a discontinuity is indicative of a phase transition.
For finite $A$, the exactly solvable model developed in
Ref.~\cite{Chase1994} leads to a rounded peak, which is shown in
Fig.~\ref{fig:m2vsm} which is a plot of $\langle m \rangle_{2}/A$ vs.
$\langle m \rangle/A$ for $\tau = 2.5$.

This behavior of the multiplicity fluctuation arises from the
sudden appearance of an infinite cluster below the critical point, a
situation that parallels a similar result in percolation models.
In percolation theory an infinite cluster exists
only when the site or bond probability exceeds a critical value as
noted above.  For
the nuclear fragmentation model, the critical condition
and the sudden presence of an infinite cluster also parallels the
discussion of Bose condensation.  Specifically, the Bose condensation
phenomenon is equivalent to a condensation of nucleons into the
largest cluster and the argument for the presence of the infinite
cluster is the same as that used to discuss the appearance of the Bose
condensed state.  Namely, since $A/x = \sum_{k} z^{k} k^{1-\tau}$ where
$x \propto V$, then for $z<1$, the sum $\sum_{k} z^{k} k^{1-\tau}$
is finite.  At $z=1$, the sum
is finite only for $\tau > 2$ and for $z >1$ the sum will diverge;
hence the sum must be truncated.  This truncation implies the
expectation of certain large clusters must be zero.  For $\tau \le 2$
an infinite cluster does not exist which is equivalent to the
condition that Bose condensation does not exist in two or fewer
dimensions.

In summary, an exactly solvable statistical model of heated hadronic
matter leads to a relationship between the critical multiplicity
$\langle m \rangle_{c}$ and critical exponent $\tau$ which
characterizes the power law
fall off of cluster yields with mass number.  The critical
multiplicity can be obtained from the peak in the multiplicity
fluctuations, which in the infinite $A$ limit has a cusp behavior due
to the sudden appearance of an infinite cluster in the theory.

This work supported in part by the National Science Foundation
Grant No. NSFPHY 92-12016.

\begin{figure}
\caption{$\langle m \rangle_{2}/A$ vs.
$\langle m \rangle/A$ for $x_{k} = x/k^{5/2}$.
The function reaches a maximum at the critical point in the infinite
$A$ limit.}
\label{fig:m2vsm}
\end{figure}


\begin{thebibliography}{99}
\bibitem{Gilkes1994} M.~L. Gilkes, et. al.,
Phys. Rev. Lett. {\bf 73}, 1590 (1994); J.~B. Elliott, et. al.
Phys. Rev. C {\bf 49}, 3185 (1994).
\bibitem{Campi1986} X. Campi,
J. Phys. A {\bf 19}, L917 (1986).
\bibitem{Chase1994} K.~C. Chase and A.~Z. Mekjian,
Phys. Rev. C {\bf 49}, 2164 (1994). A.~Z. Mekjian,
Phys. Rev. Lett. {\bf 64}, 2125 (1990); Phys. Rev. C {\bf 41}, 2103
(1990); S.~J. Lee and A.~Z. Mekjian, Phys. Rev. C {\bf 45}, 365
(1992); Phys. Rev. C {\bf 45}, 1284 (1992).
\bibitem{Finn1982} J.~E. Finn et al.,
Phys. Rev. Lett. {\bf 49}, 1321 (1982).

\end{thebibliography}
\end{document}